\documentclass[12pt,preprint]{aastex}
\usepackage{graphicx,apjfonts,emulateapj5}
\usepackage{onecolfloat5}

\shorttitle{Spatial energy spectrum of magnetic fields in our Galaxy}
\shortauthors{Han et al.}
\slugcomment{\today}

\begin{document}
\twocolumn[

\title{The spatial energy spectrum of magnetic fields in our Galaxy}
\author{J. L. Han\altaffilmark{1},
        K. Ferriere\altaffilmark{2},
	and
        R. N. Manchester\altaffilmark{3}
	}

\begin{abstract} 
Interstellar magnetic fields exist over a broad range of spatial scales,
extending from the large Galactic scales ($\sim 10$~kpc) down to the very
small dissipative scales ($\ll 1$~pc).  In this paper, we use a set of 490
pulsars distributed over roughly one third of the Galactic disk out to a
radius $R \simeq 10$~kpc (assuming $R_\odot = 8.5$~kpc) and combine their
observed rotation and dispersion measures with their estimated distances to
derive the spatial energy spectrum of the Galactic interstellar magnetic
field over the scale range $0.5 - 15$~kpc. We obtain a nearly flat spectrum,
with a 1D power-law index $\alpha=-0.37\pm0.10$ for $E_{\rm B}(k)=C\,
k^{\alpha}$ and an rms field strength of approximately $6\, \mu$G over the
relevant scales.  Our study complements the derivation of the magnetic
energy spectrum over the scale range $0.03 - 100$~pc by \citet{ms96b},
showing that the magnetic spectrum becomes flatter at larger scales.  This
observational result is discussed in the framework of current theoretical
and numerical models.
\end{abstract}

]

\altaffiltext{1}{National Astronomical Observatories, Chinese 
	Academy of Sciences, Jia 20 DaTun Road, Beijing 100012, China.
	Email: hjl@bao.ac.cn}
\altaffiltext{2}{Observatoire Midi-Pyr\'en\'ees, 31400 Toulouse, France. 
        Email: ferriere@ast.obs-mip.fr}
\altaffiltext{3}{Australia Telescope National Facility, CSIRO, PO Box 76,
        Epping, NSW 1710, Australia. Email: Dick.Manchester@csiro.au}

\section{Introduction}

On the whole, the physical properties of the interstellar gas are better
established and understood than those of the interstellar magnetic field.
The spatial power spectrum of interstellar HI has been measured using
observations of the 21-cm line in several small areas across the
sky. Emission measurements suggest a smooth power-law behavior with a 3D
spectral index $\simeq -3$ for spatial scales $\ga 5$~pc
\citep[e.g.,][]{cd83,gre93,dms+01}, while 21-cm absorption measurements
yield a 3D spectral index $\simeq -2.75$ in the scale range $0.01 - 3$~pc
\citep{ddg00}.

The spatial structure of interstellar free electrons has been probed using
observations of Galactic pulsars and extragalactic compact sources. Relying
on interstellar scintillation data, \citet{ars95} showed that the spatial
power spectrum of the interstellar free-electron density in the nearby
interstellar medium (ISM) could be approximated by a single power law with a
3D spectral index $\simeq -3.7$ (very close to the Kolmogorov value of
$-11/3$) for scales ranging from $\sim 10^{10}$~cm to $\sim 10^{15}$~cm.  By
combining their scintillation data with rotation measure (RM) fluctuation
measurements and with gradients in the average electron density, they were
able to extend the range of their observed nearly-Kolmogorov spectrum up to
a scale $\sim 10^{20}$~cm ($\sim 30$~pc).

Studies of pulsar and extragalactic radio-source RMs have yielded a wealth
of observational information on the spatial structure of the Galactic
interstellar magnetic field. However, most studies to date have focused on
the large-scale field structure \citep{sk80,hq94,rl94,id98,hmlq02} and only
a few have provided estimates for the rms amplitude and the characteristic
scale length of the turbulent magnetic field.  For instance, by adopting a
single-cell-size model for the turbulent field and analyzing the residuals
with respect to their best-fit model for the large-scale field, \citet{rk89}
obtained a turbulent field strength of $\simeq 5 \ \mu$G and a cell size of
$\simeq 55$~pc.  \citet{os93}, who did not resort to large-scale field
models, organized their pulsar data by pairs of pulsars seen in nearly the
same direction and interpreted them in the framework of a single-cell-size
model for the random field. With an assumed cell size in the range $10 -
100$~pc, they obtained a random field of amplitude $4 - 6 \ \mu$G.

As \citet{rk89} themselves acknowledged, the turbulent magnetic field cannot
be satisfactorily described by a single scale length. \citet{scs84} and
\citet{sc86} were the first authors to look into the question of how the
turbulent magnetic energy in the ISM is actually distributed in the scale
range $\sim 0.01 - 100$~pc. They used the structure function of RMs of
extragalactic sources to investigate variations in the composite quantity
$n_e \, {\bf B}$, where $n_e$ is the free-electron density and ${\bf B}$ is
the magnetic field. Their study suggests that the 3D spatial power spectrum
of fluctuations in $n_e \, {\bf B}$ at scales $\ga 0.01$~pc can be described
by a power law of index $\sim -3.1$ with an outer scale estimated at $\la
90$~pc. Later, \citet{ms96b} performed a combined analysis of the structure
functions of RM and emission measure (EM) to separate the fluctuations in
electron density and magnetic field in the scale range $\sim 0.03 -
100$~pc. They measured the RMs of 38 extragalactic sources located in a
small area of the sky near Galactic coordinates $l=144^\circ$, $b=21^\circ$,
where the EMs were deduced from measured H$\alpha$ intensities. For scale
sizes up to $\sim 4$~pc, the structure functions of RM and EM were
consistent with a 3D-turbulence model in which both electron density and
magnetic field fluctuations have Kolmogorov spectra.  Between $\sim 4$~pc
and $\sim 80$~pc, the structure functions were consistent with 2D
turbulence. \citet{ms96b} also obtained an rms amplitude of $\sim 1 \
\mu$G for the turbulent magnetic field on scales up to 4~pc.

\begin{figure*}
\includegraphics[angle=270,width=150mm]{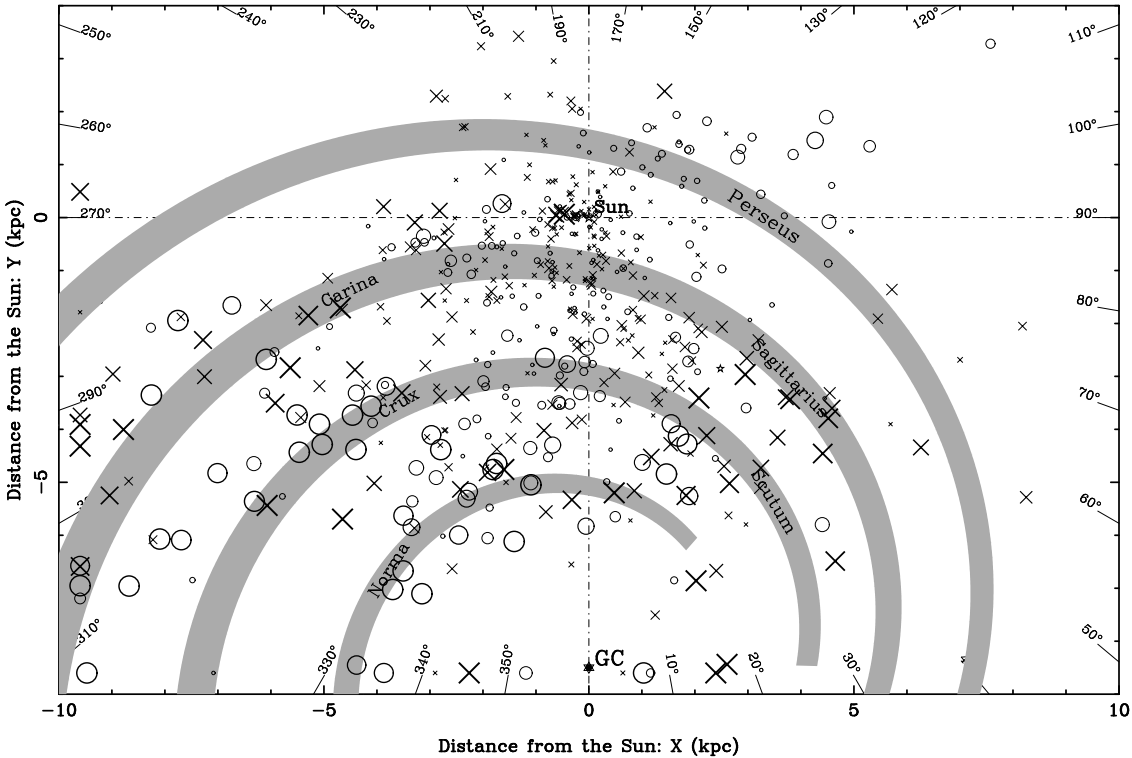}
\caption{Distribution projected on the Galactic plane of the 490 pulsars
with well-determined RMs viewed from the North Galactic pole.  Pulsars with
a positive RM are denoted with crosses, those with a negative RM are denoted
with circles, and the symbol area is proportional to the RM. Superimposed
onto the pulsar distribution is an approximate outline of the four known
spiral arms.\label{distr}}
\end{figure*}

It is clear that the magnetic energy spectrum does not abruptly stop at the
estimated outer scale of the turbulence, $\sim 80$~pc. The spectral energy
of the magnetic field will extend continuously up to the largest Galactic
scales, as a result of an inverse cascade of magnetic helicity
\citep[e.g.,][]{pfl76} and/or due to other physical processes involving
compression and shearing at large scales (e.g., Parker 1971; Kulsrud
1986). 
turbulent field and the more structured field at larger scales describe the
present-day properties of the Galactic magnetic field and provide the
necessary observational reference for theoretical models. Knowledge of the
complete magnetic energy spectrum can offer a solid observational test for
dynamo and other theories for the origin of Galactic magnetic fields.

The purpose of the present paper is to determine the portion of the magnetic
spectrum that lies above the strictly turbulent domain, thereby filling the
observational gap between the large Galactic scales and the small turbulent
scales. Pulsars are unique in providing a direct measure of the average
line-of-sight component of the interstellar magnetic field, weighted by the
local electron density, from the ratio of RM and dispersion measure (DM).
Here we will analyse the available pulsar RM and DM data in combination with
their estimated distances to deduce the interstellar magnetic energy
spectrum $E_{\rm B}$ across the scale range $0.5 - 15$~kpc. In \S~2, we
present our pulsar data and explain how they are used to construct the
interstellar magnetic energy spectrum, in \S~3, we review the assumptions
underlying our derivation, and in \S~4, we discuss the significance of our
results and compare them with theoretical predictions.

\section{The magnetic energy spectrum from pulsar data}

\subsection{Derivation of $\langle E_{\overline B_{\parallel}} \rangle (S)$} 

At the present time, we have 521 pulsars with measured RMs at our disposal:
330 of them are published measurements mostly from \citet{hl87,rl94,qmlg95}
and \citet{hmq99}, while another 202 RMs come from unpublished observations
with the Parkes telescope \citep[see ][]{hmlq02}. We discard all pulsars
with a RM uncertainty larger than 30~rad~m$^{-2}$, leaving us with a set of
490 pulsars.  The distances to the pulsars are determined from their DM
using the NE2001 model for the Galactic electron density distribution
\citep{cl03}.  The spatial distribution of our 490 pulsars is displayed in
Figure~\ref{distr}, showing that they cover roughly one third of the thick
ISM disk out to a Galactocentric radius $R \sim 10$~kpc.

The average line-of-sight component of the interstellar magnetic field
between a pulsar (P) and the Sun ($\odot$), weighted by the local
free-electron density, is given by the ratio of the pulsar's RM to its DM:
\begin{equation}
\overline{B_{\parallel}}|_{\odot}^{\rm P}
= \frac{\int_0^D n_e \ B_{\parallel} \ dl}
  {\int_0^D n_e \ dl} 
= (1.232 \ \mu {\rm G}) \ \frac{RM}{DM} \ \cdot
\label{density_aver_B}
\end{equation}
Here, $n_e$ is the free-electron density (in cm$^{-3}$), $B_{\parallel}$ is
the line-of-sight component of the magnetic field (in $\mu$G), $dl$ is the
length element along the line of sight to the pulsar (in pc), $D$ is the
distance to the pulsar (in pc), and $DM$ and $RM$ are expressed in ${\rm pc
\ cm}^{-3}$ and in ${\rm rad \ m}^{-2}$, respectively.  Likewise, the
electron-weighted average value of $B_{\parallel}$ between two pulsars,
P$_1$ and P$_2$, located on the same line of sight is given by
\begin{equation}
\overline{B_{\parallel}} |_{{\rm P}_1}^{{\rm P}_2}
= (1.232 \ \mu {\rm G}) \ \frac{RM_2 - RM_1}{DM_2 - DM_1} \ \cdot
\label{diff}
\end{equation}

\begin{figure}[t]
\includegraphics[angle=270,width=75mm]{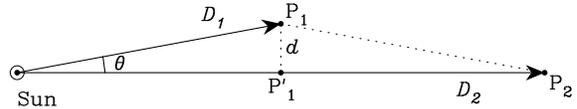}
\caption{Geometry of paired pulsars P$_1$ and P$_2$.  $\odot$ indicates the
position of the Sun and P$_1'$ represents the projection of P$_1$ onto the
line of sight to P$_2$.
\label{arro}}
\end{figure}

In order to extract as much information as possible on the spatial
variations of the interstellar magnetic field, we are going to use both
equation~(1) applied to individual pulsars and equation~(2) applied to pairs
of pulsars lying in {\it almost} the same direction.  Here, two pulsars
P$_1$ and P$_2$ are considered to lie in almost the same direction if their
angular separation, $\theta$, satisfies
\begin{equation}
d |_{{\rm P}_1}^{{\rm P}_1'} 
= D_1 \ \sin \theta  < 0.1 \  \min (D_1, \ D_2 - D_1) \ ,
\label{eq:pair}
\end{equation}
supposing that P$_1$ is the closer pulsar (see Figure~\ref{arro}).  As we
will see from equation~(\ref{key}) and Figure~\ref{yks} below,
$\overline{B_{\parallel}} |_{{\rm P}_1}^{{\rm P}_2}$ is only sensitive to
magnetic fluctuation cells with sizes $\ga 0.1 \, S$ (where $S = D_1$ or
$D_2- D_1$).  Equation~(\ref{eq:pair}) ensures that line segments $\odot
{\rm P}_1$ and $\odot {\rm P}_1'$ pass through the same magnetic fluctuation
cells with sizes $\ga 0.1 \, S$.  The maximum angular separation between two
paired pulsars is $5.7^\circ$.  Note that our pair-selection approach is
similar to that originally proposed by \citet{os93} for estimating the rms
amplitude of the random field.

\begin{figure}[t]
\includegraphics[angle=270,width=87mm]{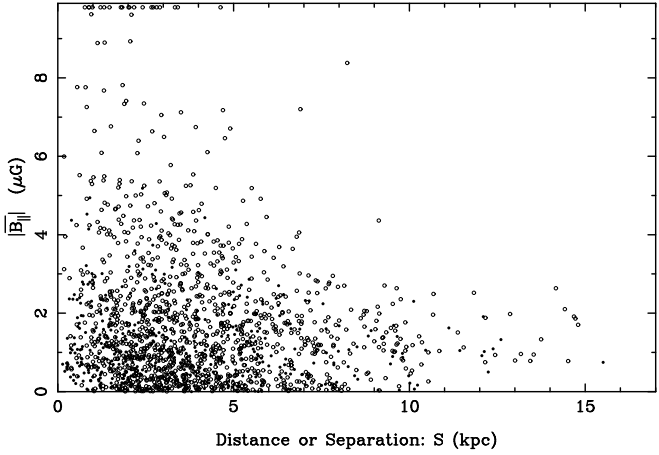} 
\caption{Measured values of $|\overline B_\parallel|$ as a function of $S$,
where $S$ is either the pulsar distance or the separation of pulsar pairs.
The 490 filled circles correspond to measurements to individual pulsars
while the 1200 open circles correspond to measurements between paired
pulsars.
\label{B_S}}
\end{figure}

Altogether, we have 490 measurements of $\overline{B_{\parallel}}$ to
individual pulsars and 1200 measurements between paired pulsars.  For the
following, we denote by $S$ the distance over which
$\overline{B_{\parallel}}$ is measured, i.e., $S = D$ for the first type of
measurements and $S = D_2- D_1$ for the second type. Measured values of
$|\overline{B_{\parallel}}|$ are shown as a function of $S$ in
Figure~\ref{B_S}. A few pulsar pairs had similar DMs but very different RMs,
resulting in anomalously high values of $\overline{B_{\parallel}}$. In fact,
the separation of these pulsars is almost certainly greater than indicated
by the DMs, so these values were capped at 15 $\mu$G. These and the few
values between 10 and 15 $\mu$G are plotted at the top of the figure.

\begin{figure*}
\includegraphics[angle=270,width=130mm]{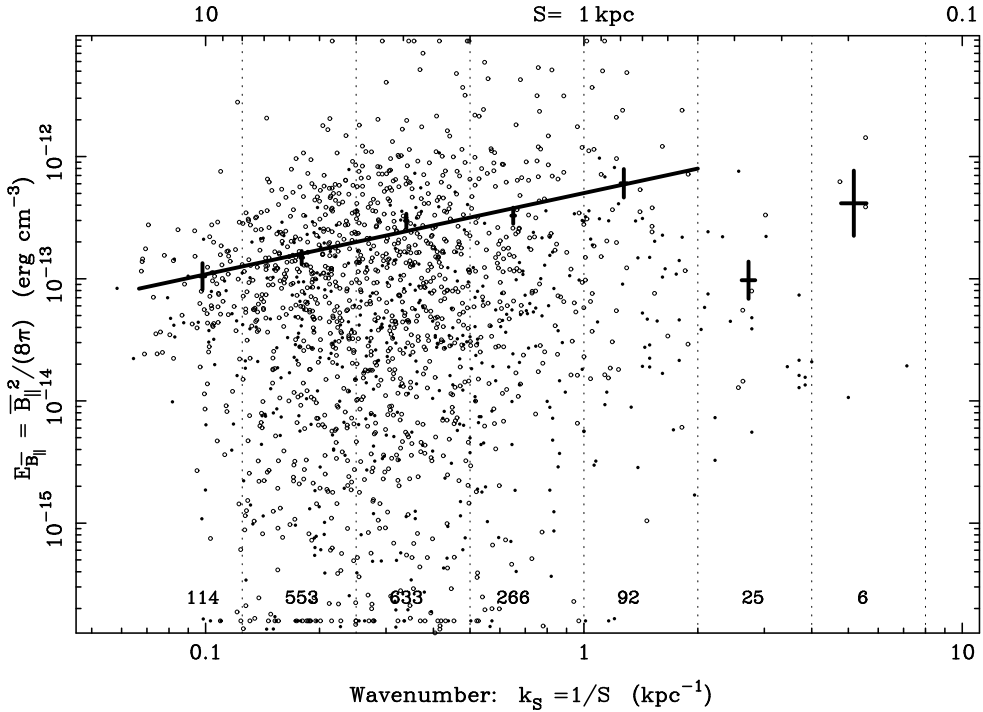} 
\caption{Measured values of the magnetic energy associated with
$\overline{B_{\parallel}}$, $E_{\overline B_{\parallel}} =
\overline{B_{\parallel}} ^2 / (8 \pi)$, versus wavenumber, $k_S = 1 / S$,
where $S$ is either the pulsar distance or the separation of pulsar pairs.
The thick crosses give average values $\langle E_{\overline B_{\parallel}}
\rangle$ and their uncertainties over the 7 successive intervals delimited
by the vertical dotted lines and corresponding, from right to left, to
$S$(kpc) = [0.125,0.25], [0.25,0.5], [0.5,1.0], [1.0,2.0], [2.0,4.0],
[4.0,8.0], [8.0,16.0].  The oblique thick line is a power-law fit to the
$\langle E_{\overline B_{\parallel}} \rangle$ with $k_S < 2$ kpc$^{-1}$ (see
eq.~[\ref{eebk}]).  The number of data points in each interval is indicated
at the bottom.
\label{ebsk}}
\end{figure*}

The magnetic energy associated with $\overline{B_{\parallel}}$ for each
pulsar or pulsar pair, $E_{\overline B_{\parallel}} \equiv
\overline{B_{\parallel}} ^2 / (8 \pi)$, is plotted as a function of $k_S
\equiv 1/S$ in Figure~\ref{ebsk} on logarithmic scales. There is a wide
scatter in the points, largely because, for any pulsar located at a given
distance $S$ and for any pair of pulsars separated by a given $S$,
$|\overline B_\parallel|$ and $E_{\overline B_{\parallel}}$ can vary from
zero to some maximum value depending on the angle between the line of sight
and the direction of the local magnetic field as well as on the position of
the pulsar or pulsar pair with respect to the maxima and minima of the
magnetic fluctuations. Never-the-less, by averaging $E_{\overline
B_{\parallel}}$ over successive intervals along the $k_S$-axis, we can
obtain estimates for the mean value of the magnetic energy associated with
$\overline{B_{\parallel}}$,
\begin{equation}
\langle E_{\overline B_{\parallel}} \rangle =
\left\langle \frac {\overline{B_{\parallel}}^2} {8 \pi} \right\rangle \ ,
\label{defE}
\end{equation}
as a function of $k_S$. It is important to realize that the above
averaging procedure (denoted by angle brackets) implies averages over
many different Galactic locations and many different lines of sight.
Despite the wide scatter in $E_{\overline B_{\parallel}}$, the average
values obtained for $\langle E_{\overline B_{\parallel}} \rangle$
(indicated with thick crosses in Figure~\ref{ebsk}) show a clear trend
with $k_S$, at least for $k_S \la 2$~kpc$^{-1}$.  
A least-squares fit of a power law to our estimated 
$\langle E_{\overline B_{\parallel}} \rangle$ gives
\begin{equation}
\langle E_{\overline B_{\parallel}} \rangle (S)
\ = \ C_0 \, \left( {k_S \over k_0} \right)^{\beta}\ , \label{eebk}
\end{equation}
with $\beta=0.66\pm0.10$, $C_0=(5.1\pm0.7) \times 10^{-13}$~erg~cm$^{-3}$,
$k_S = 1/S$, and the normalising factor $k_0 = 1$~kpc$^{-1}$.

Individual pulsar distances are uncertain by $\sim 20$\% or about
0.5~kpc at distances of a few kpc and there are only few pulsars
within 0.5~kpc from the Sun. Therefore, our estimates for the parallel
field strength and energy are unreliable at scales $\la 0.5$~kpc. 
In addition, there are only a few pulsars with distances greater than
15~kpc, so the range of validity of equation~(\ref{eebk})
is approximately $0.5\ {\rm kpc} < S < 15\ {\rm kpc}$.

\subsection{Relationship between 
$\langle E_{\overline B_{\parallel}} \rangle (S)$ and $E_{\rm B} (k)$ }

The next important step is to relate our measured
$\langle E_{\overline B_{\parallel}} \rangle (S)$ to the magnetic-energy 
spectral density, $E_{\rm B} (k)$.
We proceed on the assumption (discussed in \S~3)
that the spatial fluctuations of the magnetic field, ${\bf B}$, are 
statistically homogeneous and isotropic.
We start by writing ${\bf B} ({\bf r})$ in terms of its three-dimensional 
Fourier transform, $\tilde{\bf B} ({\bf k})$:
\begin{equation}
{\bf B} ({\bf r}) = \int \tilde{\bf B} ({\bf k}) \
e^{2\pi i \, {\bf k} \cdot {\bf r}} \ d{\bf k} \ ,
\label{bbrs}
\end{equation}
where ${\bf r}$ is the position vector and ${\bf k}$ is the wave
vector defined such that the relationship between wavenumber, $k$, and
wavelength, $\lambda$, is $k = 1 / \lambda$.  The line-of-sight
component of ${\bf B}$ at a distance $l$ can then be written as
\begin{equation}
B_{\parallel} (l) = \int \tilde{B}_{\parallel} ({\bf k}) \
e^{2\pi i \, k_{\parallel} \, l} \ d{\bf k} \ \cdot
\label{Bpar}
\end{equation}
Its average value between two points located on the same line of sight 
at distances $D_1$ and $D_2$, respectively, reads
\begin{equation}
\overline{B_{\parallel}}
= \frac{1}{D_2- D_1} \int_{D_1}^{D_2} B_{\parallel} (l) \ dl
\label{space_aver_B}
\end{equation}
or, after substitution of equation~(\ref{Bpar}) and integration over $l$,
\begin{equation}
\overline{B_{\parallel}}
= \int \tilde{B}_{\parallel} ({\bf k}) \
{\rm sinc}\, (\pi \, k_{\parallel} \, S) \
e^{2\pi i \, k_{\parallel} \, D_{\rm m}} \ d{\bf k} \ ,
\label{bsps}
\end{equation}
with $S = D_2- D_1$, $D_{\rm m} = (D_1 + D_2) / 2$,
and ${\rm sinc}\, (x) = (\sin x) / x$.

The expectation value of the magnetic energy associated with 
$\overline{B_{\parallel}}$ may be defined as
\begin{eqnarray}
\langle E_{\overline B_{\parallel}} \rangle (S)
& = &
\frac{1}{8\pi} \
\langle \overline{B_{\parallel}} \
\overline{B_{\parallel}}^{*} \rangle 
\nonumber \\
& = &
\frac{1}{8\pi} \int \int 
\langle \tilde{B}_{\parallel} ({\bf k}) \ 
\tilde{B}_{\parallel}^{*} ({\bf k}^{'}) \rangle \
{\rm sinc}\, (\pi \, k_{\parallel} \, S) \
{\rm sinc}\, (\pi \, k_{\parallel}^{'} \, S) \
\nonumber \\
& &
\phantom{ \frac{1}{8\pi} \int \int }
e^{2\pi i \, (k_{\parallel} - k_{\parallel}^{'}) \, D_{\rm m}} \ 
d{\bf k} \ d{\bf k}^{'} \ \cdot
\label{eeks1}
\end{eqnarray}
In the last line of the above equation, the expectation value is applied
to the factor $\tilde{B}_{\parallel} ({\bf k}) \
\tilde{B}_{\parallel}^{*} ({\bf k}^{'})$, which is the only stochastic
part of the expression.
Now, for homogeneous and isotropic turbulence, we have
\begin{eqnarray}
\frac{1}{8\pi} \
\langle \tilde{B}_{\parallel} ({\bf k}) \ 
\tilde{B}_{\parallel}^{*} ({\bf k}^{'}) \rangle 
& = &
\frac{1}{3} \ 
\left[ \frac{1}{8\pi} \ 
\langle \tilde{\bf B} ({\bf k}) \cdot 
\tilde{\bf B}^{*} ({\bf k}^{'}) \rangle 
\right]
\nonumber \\
& = &
\frac{1}{3} \ {\cal E}_{\rm B} ({\bf k}) \
\delta({\bf k} - {\bf k}^{'})
\nonumber \\
& = &
\frac{1}{3} \ \frac{E_{\rm B}(k)}{4\pi \, k^2} \
\delta({\bf k} - {\bf k}^{'}) \ ,
\label{correl}
\end{eqnarray}
where ${\cal E}_{\rm B} ({\bf k})$ is the 3D magnetic-energy spectral
density and $E_{\rm B}(k)$ is its 1D counterpart.
In view of equation~(\ref{correl}), equation~(\ref{eeks1}) reduces to
\begin{equation}
\langle E_{\overline B_{\parallel}} \rangle (S) =
\frac{1}{3} \int \frac{E_{\rm B}(k)}{4\pi \, k^2} \
{\rm sinc}^2 (\pi \, k_{\parallel} \, S) \ d{\bf k} \ \cdot
\label{eeks2}
\end{equation}
The integration over the direction of ${\bf k}$ is conveniently carried out 
in a cylindrical coordinate system with polar axis along the line of sight,
such that $k_{\parallel} = k \, \cos \theta$ and 
$d{\bf k} = 2\pi \, k^2 \, dk \ d(\cos \theta)$.
The final result can be cast in the form
\begin{equation}
\langle E_{\overline B_{\parallel}} \rangle (S) =
\frac{1}{3} \int_0^{\infty} E_{\rm B}(k) \ y(k \, S) \ dk \ ,
\label{key}
\end{equation}
with
\begin{equation}
y(k \, S) \equiv \frac{1}{\pi k \, S} \int_0^{\pi k \, S} 
\left( \frac{\sin x} {x}\right)^2 \ dx \ .
\label{weight}
\end{equation}
Clearly, the $1/3$ prefactor in equation~(\ref{key}) follows
from our assumption of isotropy whereby one third of the magnetic
energy goes into the line-of-sight component. 

\begin{figure}[t]
\includegraphics[height=80mm,width=55mm,angle=270]{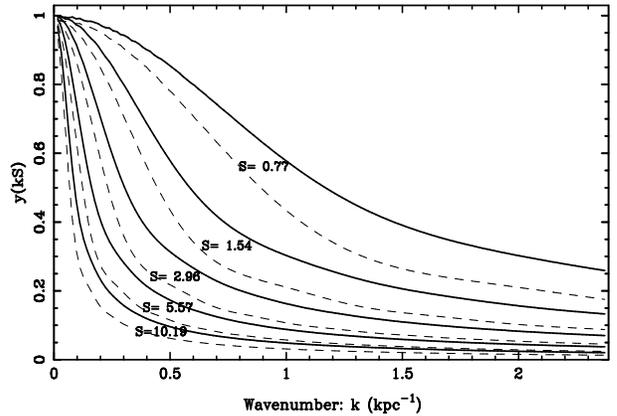}
\caption{Weight function, $y(k\,S)$ (defined by eq.~[\ref{weight}]), 
as a function of wavenumber, $k$, for 5 selected values of the pulsar
separation, $S$ (corresponding to the averaged $S$ in the first 5 
intervals of Figure~\ref{ebsk}), in the 3D case. 
The dashed lines show the weight function in the 2D case 
(eq.~[\ref{2dwf}]). 
\label{yks}}
\end{figure} 

The shape of the weight function, $y(k \, S)$, is displayed for different
values of the pulsar separation $S$ in Figure~\ref{yks}. For any given $S$,
magnetic fluctuations with $k\,S \ll 1$, i.e., with wavenumber
$k \ll k_S$ or wavelength $\lambda \gg S$, give their full weight to 
$\langle E_{\overline B_{\parallel}} \rangle (S)$.
Fluctuations with $k \sim k_S$ or $\lambda \sim S$ give approximately 
half their weight, and fluctuations with $k \gg k_S$ or $\lambda \ll S$ 
make a negligible contribution.
Hence the main contribution to $\langle E_{\overline B_{\parallel}} 
\rangle (S)$ comes from fluctuations with wavenumber $k \la k_S$. 
Physically, for long-wavelength fluctuations ($\lambda \gg S$),
$B_{\parallel}$ is nearly constant between both pulsars, so that its average
value ${\overline B_{\parallel}}$ is approximately equal to its local value.
For short-wavelength fluctuations ($\lambda \ll S$), $B_{\parallel}$
reverses many times between both pulsars, so that the contributions to
$\overline{B_{\parallel}}$ from line segments with opposite sign of
$B_{\parallel}$ cancel each other out almost entirely. For fluctuations with
$\lambda = S$, there is complete cancellation if ${\bf k}$ is along the line
of sight, and there is no cancellation if ${\bf k}$ is perpendicular to the
line of sight; an average over the direction of ${\bf k}$ then leads to a
net weight factor $\sim 0.5$.

\subsection{Determination of $E_{\rm B}(k)$} 

In practice, we tried various analytical functions for $E_{\rm B}(k)$.  For
each of them, we computed the corresponding function $\langle E_{\overline
B_{\parallel}} \rangle (S)$ by integrating the right-hand side of
equation~(\ref{key}) numerically and then compared the computed $\langle
E_{\overline B_{\parallel}} \rangle$ to the measured $\langle E_{\overline
B_{\parallel}} \rangle$ displayed in Figure~\ref{ebsk}.  The lower limit of
integration as well as the step length of $k$ were set to 0.02~kpc$^{-1}$.
The upper limit of integration is mainly determined by the shape 
of the weight function, $y(k \, S)$, shown in Figure~\ref{yks};
here we adopted the value at which the function
$E_{\rm B}(k) \, y(k \, S)$ drops below 0.1\% of its peak value.

The trial function $E_{\rm B}(k)$ that provides the best fit to the measured
$\langle E_{\overline B_{\parallel}} \rangle$ is a slowly decreasing power
law described by
\begin{equation}
E_{\rm B}(k) = C \, \left( {k \over k_0} \right)^{\alpha} \ , \label{spectrum}
\end{equation}
with $\alpha=-0.37\pm0.10$ and $C = (6.8\pm0.8) \times
10^{-13}$~erg~cm$^{-3}$~kpc.  As a reminder, $k$ is the wavenumber defined
as $1/\lambda$ and $k_0 = 1$~kpc$^{-1}$.  The power-law indices of $\langle
E_{\overline B_{\parallel}} \rangle (S)$ (see eq.~[\ref{eebk}]) and $E_{\rm
B}(k)$ (see eq.~[\ref{spectrum}]) are approximately related through $\beta
=1+\alpha$.  This relation would be exact if the weight function $y(k \, S)$
were approximated by the step function $u(1 - k \, S)$, where $u(x)$ is
defined such that $u(x) = 0$ for $x < 0$ and $u(x) = 1$ for $x >
0$. Consequently, the uncertainty in $\alpha$ is comparable to the
uncertainty in $\beta$, i.e., $\simeq 0.10$.  Furthermore, since $\langle
E_{\overline B_{\parallel}} \rangle (S)$ has been reliably determined over
the range $0.5\ {\rm kpc} < S < 15\ {\rm kpc}$ (see Section~2.1), the
validity range of equation~(\ref{spectrum}) is approximately $0.07\ {\rm
kpc}^{-1} < k < 2\ {\rm kpc}^{-1}$.

The rms value of the fluctuating magnetic field with wavenumbers in the
above range, obtained by integrating equation~(\ref{spectrum}), is $6.1\pm
0.5 \ \mu$G. We note that this rms field strength is consistent with
estimates by previous authors, e.g., via modelling diffuse $\gamma$-ray data
and synchrotron emission \citep{smr00}, via energy equipartition
(Berkhuijsen, referenced in Beck et al. 1996)\nocite{bbm+96}, or based on
pulsar data and synchrotron data by \citet{hei96b}. This leads us to believe
that equation (1) using pulsar RM and DM data does not significantly
underestimate or overestimate the field strength in the ISM, at least on
average over a large sample.

\section{Validity of the underlying assumptions}

We have used the tools of homogeneous and isotropic turbulence theory
to derive the spatial power spectrum of magnetic energy in our
Galaxy. However, strictly speaking, these are not directly applicable
to the problem at hand. Firstly, the spatial variations
of the magnetic field at the scales probed here are not all
stochastic; some are associated with large-scale coherent features such as
supernova remnants or superbubbles and others are associated
with the large-scale Galactic spiral structure.  In this study, we did
not try to distinguish the purely stochastic fluctuations from the
coherent variations; all are included in our
magnetic energy spectrum.  This procedure is justified if the results
are taken to represent present-day Galactic averages
rather than true statistical averages.  

A second, more fundamental,
reason is that the magnetic energy distribution is neither homogeneous
across the sampled region (it includes spatial variations
up to scales approaching the size of the region, which is itself a
substantial portion of the Galaxy) nor isotropic (the effects of the
disk geometry and the associated vertical stratification are
significant at the relevant spatial scales).  The most obvious
manifestation of the lack of isotropy is the dominant horizontal
component of the large-scale magnetic field.

The failure of the assumption of homogeneity is inherent in our scale
range.  \citet{ms96b} did not encounter the same problem because they
studied only small-scale fluctuations, such that they were able to
perform statistically meaningful spatial averages over a region which
is small enough for turbulence to be assumed homogeneous across it.
The downside is that their spectrum applies only to their small region
of observations and to small scales. In contrast, our derived spectrum
does not constitute a good statistical average.  However, since our
pulsar pairs are reasonably well distributed throughout the sampled
volume, our spectrum may be regarded as a spatial average over the
sampled volume -- with some bias toward the vicinity of the Sun. Since
the sampled volume corresponds roughly to a $120^\circ$ wedge of the
Galactic disk seen from the Galactic center, it is probably
representative of the whole disk. In this case, our derived spectrum
also provides a good estimate for the Galaxy-wide average spectrum.

The assumption of isotropy is almost unavoidable and routinely made in
practice \citep[e.g.][]{os93,ms96b}, even though it fails both at large
scales (for the reason outlined above) and at small scales (due to the
presence of the large-scale magnetic field). Here, the use of the isotropy
assumption and of the ensuing formulae would be justified if the sky
coverage in line-of-sight direction were complete and uniform and if the
magnetic spectrum were homogeneous (so as to give all directions the same
weight).  However, neither condition is fully satisfied.  The resulting
uncertainty can be estimated once it is realized that the situation we are
dealing with is intermediate between 3D isotropic and 2D isotropic (with the
magnetic field at all locations and all sightlines being parallel to the
Galactic plane).  The 2D case can be treated similarly to the 3D case
studied in \S~2.2 and \S~2.3: the expression found for $\langle E_{\overline
B_{\parallel}} \rangle (S)$ is identical to equation~(\ref{key}) with the
$1/3$ prefactor replaced by $1/2$ and the weight function given by
\begin{equation}
y(k \, S) \equiv \frac{2}{\pi}
\int_0^{\pi k \, S} \frac{1}{\sqrt{(\pi k \, S)^2 -
x^2}} \ \left(\frac{\sin x} {x}\right)^2 \ dx \ .
\label{2dwf}
\end{equation}
A numerical fit to Figure~\ref{ebsk} then leads to a corresponding
equation~(\ref{spectrum}) with $\alpha = -0.39\pm 0.10$ and $C_{\rm 2D} =
(8.6 \pm 1.0) \times 10^{-13}$ erg cm$^{-3}$~kpc.  This shows that the
derived values are not very sensitive to the isotropy assumption.  The
reason why a similar spectral index is obtained in the 2D and 3D cases is
easily understood.  In \S~2.3, we saw that the spectral index obtained with
the step-function approximation to the weight function ($\alpha = \beta - 1
= -0.34$) is close to that obtained with the correct weight function
($\alpha = -0.37$).  This indicates that the slope of the energy spectrum
depends little on the exact shape of the weight function, but is governed by
the relative positions of the cutoff wavenumbers.

Another important assumption implicit in our derivation is the absence of
correlation between the fluctuations in electron density and those in
magnetic field strength, which allows us to approximate the space-averaged
value of $B_{\parallel}$ between two points (eq.~[\ref{space_aver_B}]) by a
density-weighted average (eq.~[\ref{density_aver_B}] or [\ref{diff}]).  This
assumption may not be well satisfied in the ISM. Both positive and negative
correlations are possible and their net effect is difficult to quantify
\citep{bssw03}. However, as mentioned in \S~2.3, our derived rms field
strength is consistent with other estimates. In addition, the fluctuation
spectral index is affected only if the degree of correlation is
scale-dependent.

Finally, individual pulsar distances are uncertain by a large factor,
typically $\sim 20\%$, although the overall distance scale is more
accurate. The impact on our derived magnetic spectrum is limited by
the fact that the values of $\langle E_{\overline
B_{\parallel}}\rangle (S)$ were obtained from the average of many
measurements of $\overline{B_{\parallel}}^2$.

\section{Discussion}

We have derived the interstellar magnetic energy spectrum over the
scale range $0.5 - 15$~kpc, based directly on measurements of pulsar
RMs and DMs. The outer scale is a significant fraction of the size of
the Galaxy and so contributions from the so-called large-scale (or
regular) magnetic field are included. The spectral region covered by
our derived spectrum has never been studied observationally before,
yet it constitutes a crucial piece of the complete spectrum, whose
knowledge is important for our understanding of the behavior and
evolution of Galactic magnetic fields.

\begin{figure}[t]
\includegraphics[angle=270,width=87mm]{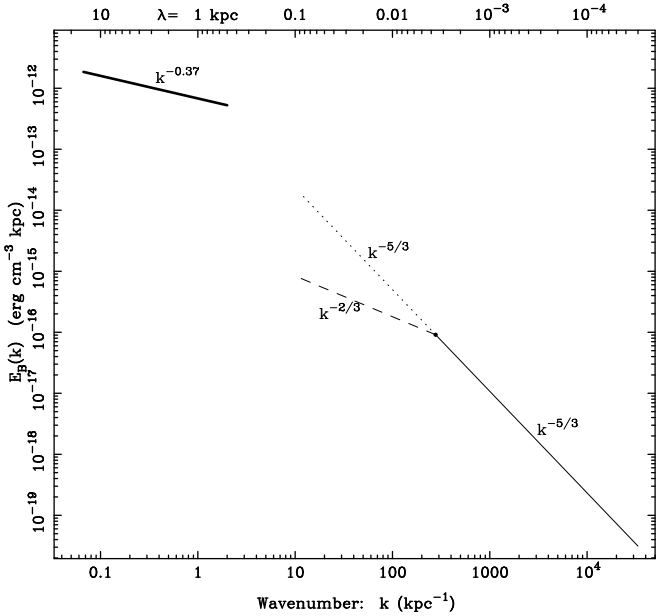}
\caption{Composite magnetic energy spectrum in our Galaxy.  
The thick solid line is the large-scale spectrum derived in this paper.  
The thin solid and dashed/dotted lines give the Kolmogorov and
2D-turbulence spectra, respectively, inferred from the \citet{ms96b} study. 
Their Kolmogorov spectrum, after adjusting to our definition of $k$, 
can be written as $E_{\rm B}(k) = C_{\rm K} \, (k/k_0)^{-5/3}$ with
$C_{\rm K} = 9.5 \times 10^{-13}$~erg~cm$^{-3}$~kpc and $k_0=1$~kpc$^{-1}$.
The 2D-turbulence spectrum is uncertain (see main text);
it probably lies between the dashed ($E_{\rm B}(k) \propto k^{-2/3}$) and 
dotted ($E_{\rm B}(k) \propto k^{-5/3}$) lines, which are both scaled
to match the Kolmogorov spectrum at $k_{\rm 3D} = (4~{\rm pc})^{-1}$.
\label{gs}}
\end{figure}

In Figure~\ref{gs}, we plot our measured spectrum with that inferred
from the \citet{ms96b} study at smaller scales.  This spectrum has a
Kolmogorov slope ($E_{\rm B}(k) \propto k^{-5/3}$) up to a scale
$l_{\rm 3D} \sim 4$~pc and an uncertain shape consistent with 2D
turbulence between $l_{\rm 3D}$ and $l_{\rm 2D} \sim 80$~pc. If, as
suggested by \citet{ms96b}, the 2D turbulence is in the form of thin
sheets with thickness $\sim l_{\rm 3D}$, we could assume that
fluctuations with scales larger than $l_{\rm 3D}$ persist only if
their projected wavelength normal to a sheet, $\lambda \ \cos \theta$,
is less than the sheet's thickness $l_{\rm 3D}$. The fraction of
acceptable wave vectors at a given wavenumber $k$ is then $l_{\rm 3D}
/ \lambda = k / k_{\rm 3D}$. If the acceptable wave vectors preserve a
Kolmogorov spectral energy, the magnetic spectrum in the wavenumber
range $k_{\rm 2D} - k_{\rm 3D}$ is simply the Kolmogorov spectrum
multiplied by the above factor, i.e., a $k^{-2/3}$ spectrum.  On the
other hand, if 2D turbulence means that the spectral energy is simply
redistributed without loss into two dimensions,
then the Kolmogorov spectrum $\propto k^{-5/3}$ extends out to $k_{\rm
2D}$.  The actual situation probably lies between these two extremes.

The two important features emerging from the composite spectrum
(Figure~\ref{gs}) are (1) a change in slope from $\alpha = -5/3$ at
the small ($\la 4$~pc) scales to $\alpha \simeq -0.37$ at the large
($\ga 0.5$~kpc) scales and (2) a possibly significant discontinuity
between $\sim 80$~pc and $\sim 0.5$~kpc.  How do these two features
fit with theoretical predictions?

According to our theoretical understanding of the ISM, turbulent energy is
injected into the ISM by stellar winds and supernova explosions on scales
$\sim 10 - 100$~pc.  Magnetic energy is then transferred, via a so-called direct
cascade, toward smaller and smaller scales. In addition, because
interstellar turbulence is helical, there is also an inverse cascade of
magnetic helicity toward larger and larger scales, making it possible to
amplify or maintain a large-scale magnetic field \citep{pfl76}.  Because
the quantity that is transferred as a whole in the domain of the inverse
cascade is magnetic helicity ($H_{\rm B}(k) \propto E_{\rm B}(k) / k$)
rather than magnetic energy, the corresponding spectrum is flatter than in
the domain of the direct cascade.  Therefore, our finding a flatter spectrum
at the larger scales is consistent with the inverse cascade theory. 

However, the value obtained for the spectral index ($\alpha \simeq -0.37$)
differs from current theoretical estimates.  In the theory of 3D MHD helical
turbulence developed by \citet{pfl76}, a numerical solution of the evolution
equation for the magnetic energy spectrum yields $E_{\rm B}(k) \propto
k^{-1}$ at scales larger than the injection scale. As shown by these
authors, the same spectral slope is obtained by applying a dimensional
Kolmogorov-type argument to the inverse cascade of magnetic helicity.
Direct numerical simulations do not yet convey an unambiguous or consistent
picture.  \citet{bp99} obtained numerical spectra consistent with the
$k^{-1}$ theoretical prediction.  In contrast, in the simulations of
\citet{bra01}, a wave of spectral magnetic energy propagates toward smaller
$k$ under a $k^{-1}$ envelope, but no $k^{-1}$ spectrum actually develops.
Instead, magnetic energy builds up at the largest scales and a secondary
peak appears at the injection scale. \citet{mb02} showed that the actual
shape of the magnetic energy spectrum depends in fact quite sensitively on
the level of fractional kinetic helicity with which the turbulence is
forced. According to their simulations, high levels of kinetic helicity lead
to double-peaked spectra similar to that obtained by \citet{bra01}, whereas
low levels of kinetic helicity give rise to a single peak at the small
resistive scale.  It is clear that one has to be extremely cautious when
trying to apply numerical spectra to the real Galactic magnetic field.
Current simulations are still highly idealized (e.g., uniform background
medium, incompressible or isothermal gas, periodic boundary conditions), and
they neglect potentially important factors such as the true nature of
turbulent forcing, the presence of large-scale shear, and the actual
boundaries of the Galactic disk.  Another difficulty is that the saturated
spectra obtained numerically emerge after resistivity comes into play,
whereas the Galaxy is probably not yet in the resistive regime (Blackman,
private communication).

The discontinuity appearing between the <100pc and >0.5kpc scales in
the composite spectrum of Figure~\ref{gs} can be partly explained by the
fact that the different parts of the spectrum apply to different Galactic
regions.  As discussed in \S~3, our portion of the spectrum was obtained
from a magnetic energy average over about one third of the inner Galactic
disk, while that of \citet{ms96b} was obtained from a small high-latitude
region in the outer Galaxy, where the fields are presumably much weaker than
in the inner disk.  However, if the 2D-turbulence portion of the spectrum is
as flat as $k^{-2/3}$, the observed discontinuity is so severe that it
suggests a genuine excess in the Galactic magnetic energy spectrum at large
scales. This excess compared to the amount expected from the inverse cascade
alone could indicate that most of the energy input to the large-scale
magnetic field occurs directly at large scales.  This interpretation would
be consistent with the standard Galactic dynamo theory, in which the
large-scale magnetic field is amplified or maintained through the combined
action of small-scale helical turbulence, presumably via the inverse cascade
leading to the so-called alpha-effect, and large-scale shear, typically
associated with the Galactic differential rotation
\citep[e.g.,][]{par71,vr71}. It is widely believed that the large-scale
shear acts much more efficiently than the helical turbulence
\citep[e.g.,][]{par71,rss88,bdm+92,fs00,bla00}, especially when the spiral
structure of the Galaxy and the streaming motions along spiral arms are
taken into account \citep[e.g.,][]{rre99,eovu00}. A more extreme point of
view would be to consider that the large-scale magnetic field is not
affected at all by the small-scale turbulent field and that all the energy
contained in the large-scale field is directly injected at large scales.
Various scenarios have been proposed in this spirit, such as protogalactic
collapse and subsequent shearing by the Galactic differential rotation with
rapid ambipolar diffusion \citep{kul86b} or protogalactic collapse and
subsequent excitation of spiral arms and bars resulting in nonaxisymmetric,
not-perfectly-azimuthal motions \citep{lc97}.

\section*{Acknowledgments}
We thank an anonymous referee for an extensive report which helped us
improve the clarity of the paper. We also thank Sun Xiaohui, You
Xiaopeng and Xu Jianwen for helpful discussions. JLH is supported by
the National Natural Science Foundation (NNSF) of China (10025313) and
the National Key Basic Research Science Foundation of China
(G19990754) as well as the partner group of MPIfR at NAOC. The Parkes
radio telescope is part of the Australia Telescope which is funded by
the Commonwealth Government for operation as a National Facility
managed by CSIRO.

\bibliographystyle{apj}
\bibliography{journals,psrrefs,modrefs}

\begin{thebibliography}{37}
\expandafter\ifx\csname natexlab\endcsname\relax\def\natexlab#1{#1}\fi

\bibitem[{Armstrong {et~al.}(1995)Armstrong, Rickett, \& Spangler}]{ars95}
Armstrong, J.~W., Rickett, B.~J., \& Spangler, S.~R. 1995, ApJ, 443, 209

\bibitem[{{Balsara} \& {Pouquet}(1999)}]{bp99}
{Balsara}, D. \& {Pouquet}, A. 1999, Physics of Plasmas, 6, 89

\bibitem[{{Beck} {et~al.}(1996){Beck}, {Brandenburg}, {Moss}, {Shukurov}, \&
  {Sokoloff}}]{bbm+96}
{Beck}, R., {Brandenburg}, A., {Moss}, D., {Shukurov}, A., \& {Sokoloff}, D.
  1996, ARA\&A, 34, 155

\bibitem[{Beck {et~al.}(2003)Beck, Shukurov, Sokoloff, \& Wielebinski}]{bssw03}
Beck, R., Shukurov, A., Sokoloff, D., \& Wielebinski, R. 2003, A\&A, in press.

\bibitem[{{Blackman}(2000)}]{bla00}
{Blackman}, E.~G. 2000, ApJ, 529, 138

\bibitem[{{Brandenburg}(2001)}]{bra01}
{Brandenburg}, A. 2001, ApJ, 550, 824

\bibitem[{Brandenburg {et~al.}(1992)Brandenburg, Donner, Moss, Shukurov,
  Sokoloff, \& Tuominen}]{bdm+92}
Brandenburg, A., Donner, K.~J., Moss, D., Shukurov, A., Sokoloff, D.~D., \&
  Tuominen, I. 1992, A\&A, 259, 453

\bibitem[{{Cordes} \& {Lazio}(2003)}]{cl03}
{Cordes}, J.~M. \& {Lazio}, T.~J.~W. 2003, ApJ, submitted

\bibitem[{{Crovisier} \& {Dickey}(1983)}]{cd83}
{Crovisier}, J. \& {Dickey}, J.~M. 1983, A\&A, 122, 282

\bibitem[{{Deshpande} {et~al.}(2000){Deshpande}, {Dwarakanath}, \&
  {Goss}}]{ddg00}
{Deshpande}, A.~A., {Dwarakanath}, K.~S., \& {Goss}, W.~M. 2000, ApJ, 543, 227

\bibitem[{Dickey {et~al.}(2001)Dickey, McClure-Griffiths, Stanimirovi\'{c},
  Gaensler, \& Green}]{dms+01}
Dickey, J.~M., McClure-Griffiths, N.~M., Stanimirovi\'{c}, S., Gaensler, B.~M.,
  \& Green, A.~J. 2001, ApJ, 561, 264

\bibitem[{{Elstner} {et~al.}(2000){Elstner}, {Otmianowska-Mazur}, {von Linden},
  \& {Urbanik}}]{eovu00}
{Elstner}, D., {Otmianowska-Mazur}, K., {von Linden}, S., \& {Urbanik}, M.
  2000, A\&A, 357, 129

\bibitem[{{Ferri{\` e}re} \& {Schmitt}(2000)}]{fs00}
{Ferri{\` e}re}, K. \& {Schmitt}, D. 2000, A\&A, 358, 125

\bibitem[{{Green}(1993)}]{gre93}
{Green}, D.~A. 1993, MNRAS, 262, 327

\bibitem[{{Hamilton} \& {Lyne}(1987)}]{hl87}
{Hamilton}, P.~A. \& {Lyne}, A.~G. 1987, MNRAS, 224, 1073

\bibitem[{Han {et~al.}(2002)Han, Manchester, Lyne, \& Qiao}]{hmlq02}
Han, J.~L., Manchester, R.~N., Lyne, A.~G., \& Qiao, G.~J. 2002, ApJ, 570, L17

\bibitem[{{Han} {et~al.}(1999){Han}, {Manchester}, \& {Qiao}}]{hmq99}
{Han}, J.~L., {Manchester}, R.~N., \& {Qiao}, G.~J. 1999, MNRAS, 306, 371

\bibitem[{{Han} \& {Qiao}(1994)}]{hq94}
{Han}, J.~L. \& {Qiao}, G.~J. 1994, A\&A, 288, 759

\bibitem[{{Heiles}(1996)}]{hei96b}
{Heiles}, C. 1996, in ASP Conf. Ser. 97: Polarimetry of the Interstellar
  Medium, 457

\bibitem[{{Indrani} \& {Deshpande}(1998)}]{id98}
{Indrani}, C. \& {Deshpande}, A.~A. 1998, New Astronomy, 4, 33

\bibitem[{{Kulsrud}(1986)}]{kul86b}
{Kulsrud}, R. 1986, in Plasma Astrophysics, ed. T.~Guyenne \& L.~Zeleny (ESA
  Publ. SP-251, Paris), 531--537

\bibitem[{Lesch \& Chiba(1997)}]{lc97}
Lesch, H. \& Chiba, M. 1997, Fundamentals of Cosmic Physics, 18, 273

\bibitem[{{Maron} \& {Blackman}(2002)}]{mb02}
{Maron}, J. \& {Blackman}, E.~G. 2002, ApJ, 566, L41

\bibitem[{Minter \& Spangler(1996)}]{ms96b}
Minter, A.~H. \& Spangler, S.~R. 1996, ApJ, 458, 194

\bibitem[{{Ohno} \& {Shibata}(1993)}]{os93}
{Ohno}, H. \& {Shibata}, S. 1993, MNRAS, 262, 953

\bibitem[{{Parker}(1971)}]{par71}
{Parker}, E.~N. 1971, ApJ, 163, 255

\bibitem[{{Pouquet} {et~al.}(1976){Pouquet}, {Frisch}, \& {Leorat}}]{pfl76}
{Pouquet}, A., {Frisch}, U., \& {Leorat}, J. 1976, Journal of Fluid Mechanics,
  77, 321

\bibitem[{Qiao {et~al.}(1995)Qiao, Manchester, Lyne, \& Gould}]{qmlg95}
Qiao, G.~J., Manchester, R.~N., Lyne, A.~G., \& Gould, D.~M. 1995, MNRAS, 274,
  572

\bibitem[{Rand \& Kulkarni(1989)}]{rk89}
Rand, R.~J. \& Kulkarni, S.~R. 1989, ApJ, 343, 760

\bibitem[{{Rand} \& {Lyne}(1994)}]{rl94}
{Rand}, R.~J. \& {Lyne}, A.~G. 1994, MNRAS, 268, 497

\bibitem[{{Rohde} {et~al.}(1999){Rohde}, {R{\" u}diger}, \& {Elstner}}]{rre99}
{Rohde}, R., {R{\" u}diger}, G., \& {Elstner}, D. 1999, A\&A, 347, 860

\bibitem[{{Ruzmaikin} {et~al.}(1988){Ruzmaikin}, {Sokolov}, \&
  {Shukurov}}]{rss88}
{Ruzmaikin}, A.~A., {Sokolov}, D.~D., \& {Shukurov}, A.~M. 1988, {Magnetic
  fields of galaxies} (Kluwer Academic)

\bibitem[{Simard-Normandin \& Kronberg(1980)}]{sk80}
Simard-Normandin, M. \& Kronberg, P.~P. 1980, ApJ, 242, 74

\bibitem[{Simonetti \& Cordes(1986)}]{sc86}
Simonetti, J.~H. \& Cordes, J.~M. 1986, ApJ, 310, 160

\bibitem[{{Simonetti} {et~al.}(1984){Simonetti}, {Cordes}, \&
  {Spangler}}]{scs84}
{Simonetti}, J.~H., {Cordes}, J.~M., \& {Spangler}, S.~R. 1984, ApJ, 284, 126

\bibitem[{{Strong} {et~al.}(2000){Strong}, {Moskalenko}, \& {Reimer}}]{smr00}
{Strong}, A.~W., {Moskalenko}, I.~V., \& {Reimer}, O. 2000, ApJ, 537, 763

\bibitem[{{Vainshtein} \& {Ruzmaikin}(1971)}]{vr71}
{Vainshtein}, S.~I. \& {Ruzmaikin}, A.~A. 1971, Astron. Zh., 48, 902, [Sov.
  Astron., 15, 714 (1972)]

\end{thebibliography}

\end{document}